# *Press F1 for help: participatory design for dealing with on-line and real life security of older adults*

Bartłomiej Balcerzak, Wiesław Kopeć, Radosław Nielek, Sebastian Kruk,
Kamil Warpechowski, Mateusz Wasik and Marek Węgrzyn
Polish-Japanese Institute of Information Technology,
ul. Koszykowa 86, 02-008 Warsaw, Poland
Email: b.balcerzak@pjwstk.edu.pl

*Abstract*—In this paper we present the report on the design and development of a platform for the inter-generational exchange of favors. This platform was designed using participatory design approach during a 24-hour hackathon by a team consisting of younger programmers and older adults. The findings of this report show that inter-generational cooperation in which the older adults serve as representatives of the end user, not only improves the design and development of the application, but also provides an effective method for designing and applying solutions aimed at improving the security of older adults while using online and mobile tools.

## I. INTRODUCTION

Paper presents a process of designing and implementing an on-line platform dedicated for matching people that need help with volunteers wiling do offer it. Platform is intended to be used mostly, but not exclusively, by older adults. Therefore, design process has to include both older adults and young volunteers, and the final product has to meet their, often contradictory, needs. Special focus was put on designing security measures for older adults, in order to cope with their needs in this regard. Platform described in this paper was developed during 24-hours long hackathon organized in Warsaw by the Polish-Japanese Academy of Information Technology. Team developing this platform was composed of four male students and two older adults. In this case study, we aim at describing how the use of participatory design, and intergenerational communication may improve the design process of security within applications created for social interaction.

Aging is a gradual process, which is usually connected with deterioration in autonomy. House cleaning, shopping, maintenance or visiting family graves, those activities are often beyond the abilities of older adults. Some of them are done by social workers, but an opportunity to outsource part of them to volunteers might improve the coherence of society by fostering intergenerational interactions. Moreover, it might also limit the costs of social services and remove pressure from professional caregivers.

Although the most obvious scenario is the one including young volunteers helping older adults, it is not the only possibility. Surprisingly, older adults often do some volunteering works e.g. helping young kids in homework or guiding tours in local museums. It is also quite common that older adults help other people their age. Therefore, it is important to design a system that will be taking into account a variety of scenarios and will not impose predefined roles associated with age. It is also important that such system addresses the issues of safety of both the parties involved.

Demographic models indicate that in 2050 the population aged 65+ in the USA will almost double and reach 83.7 million (over 20% of the society) [11]. Report prepared by OCED [10] shows that Japan will be the oldest society in 2050. Spijker et al. argument that, although demographic data is true, in reality, the number of older adults requiring assistance in the UK and other countries has actually been falling in recent years [17].

The paper is organized as follows. Section II [Related works] describes the related works concerning volunteering of older adults, platforms for managing volunteers as well as the basics of participatory design approach. Section III [Project phases] relates to the process of the preparing and designing the platform's project. The next section, i.e. section IV [Design], describes in details the platform for exchanging favors. Final section contains conclusions and future work.

## II. RELATED WORKS

### A. Platforms for managing volunteers

Platforms that manage voluntary work are abundant on the Internet, including large communities of contributors, such as the Wikipedia [21,22]. However, virtual online platforms lack the ability of organizing non-virtual voluntary work. Platforms for managing volunteers have been so far focused on organizing the work of a homogenous group of people interested in volunteering. Products available online are mostly commercial projects that allow volunteering organizations to improve their work when in comes to administration of human resources. Prime examples of such commercially available platforms include http://ivolunteer.com or http://www.volunteerhub.com. These platforms offer mostly methods for improving the communication between the organization and the volunteers, mostly through volunteers tracking and providing methods of individual volunteer review. To the best of the authors knowledge no platform system was designed with inter-generational communication in mind.

*B. Older adults and on-line security*

According to the studies done by [4] older adults are, on average, less knowledgeable and aware of on-line risks than younger adults. Also, as shown by [1], older adults show limited trust to on-line technology. This effect, however, diminishes with the usage of ICT. Another dimension crucial for on-line security is trust. Studies done by [7] as well as [4] show that trust among older adults is not affected by age, and relies more on experience that comes with the use of ICT. Therefore, the design of applications addressing the issue of trust is important.

*C. User participation and interaction*

Fundamental idea relevant to this case is connected with the concept of user-center design, as a part of general idea of human focused approach related to another important idea: participatory design, sometimes called co-design. While there are different origins for the two latter terms, they actually refer to the same idea of bottom-up approach, which is widely used besides software engineering, from architecture and landscape design to healthcare industry.[18] All those concepts put human in the center of designing process, there is, however, a small, but significant distinction between user-center design and participatory design: the former refers the process of designing FOR users, while the latter WITH users.[14], [16]

Another case is the work done by [19], describing the cooperation between seniors and preschool children in a design task. An interesting observation made in this research was the fact that both groups needed the equal amount of time together and in separation in order to function properly. The broader context for these topics is covered by the contact theory, a widely recognized psychological concept coined by Allport [2] and developed for many years by others, e.g. Pettigrew.[12] According to the theory, the problem of intergroup stereotypes can be faced by intergroup contact. There are several conditions for optimal intergroup contact, but many studies proved that intergroup contact is a worthwhile approach, since it typically reduces intergroup prejudice.[13]

*D. Volunteering of older adults*

The effects that volunteering has on older adults is a developing field of study within various disciplines. The consensus that can be reached throughout various studies, is that participation in volunteer activity has many positive effects on the elderly. [6] claim that older adults who frequently volunteer in various activities, tend to have improved physical and mental health, compared to those who do not participate in volunteering. The work of [9] extends this correlation to wellbeing (with volunteering being correlated with higher levels of well-being), this is also reinforced by [3]and [5]. This is crucial since studies also show that in some regards, elders are more likely to be engaged in volunteer activity [8].

III. PROJECT PHASES

According to Sanders [15] there is the distinction between main product design process and pre-design stage. The latter is an early stage usually related to idea development process. In case described in this paper the pattern was observed: the design process was preceded with pre-design phase with clearly visible idea development stage.

*A. Pre-design stage*

In pre-design stage there was a vivid and intense process of idea development between both parties: the younger programmers (four male undergraduate students of the Polish-Japanese Academy of Information Technology) and senior team members (a woman and a man – a pair of participants of the Living Laboratory PJAIT, distributed laboratory and platform devoted to IT solutions development in cooperation with the city of Warsaw, Poland).

The developers team had their own idea of application, which was then discussed with elder team members. During team brainstorming process in pre-design stage final solution emerged: an on-line platform for matching older adults with younger volunteers in order to exchange services.

*B. Design stage*

During the design stage the insight provided by the seniors working with the team was applied to the construction of the platform. Upon discussion between members of the two generations, the name 'F1' was chosen for the application. The name, inspired by the F1 hotkey used in various kinds of software to open the help window, was proposed by one of the seniors, who knew of the hotkey's role from previous experience and computer courses he had participated in. Based on the discussion made during the pre-design stage and the topics raised by the older adults, focus was put by the programmers on designing solutions related to accessibility of the platform for all users, and issues of user safety and security of the exchange of favors. During the design stage the role of seniors was limited mostly due to the lack of technical expertise required in the process of designing a platform. This does not mean, however, that the older adults were passive. They tried to be active and engaged throughout the whole process e.g. via e-mail or phone. As it was mentioned above, participatory design approach can be applied at any stage of the process, but its influence is the most apparent and significant in pre-design stage. We observed a similar shift in time: from co-design at early stage of teamwork to more typical user-center approach towards the end of the hackathon. According to the workflow observed, the opinions that seniors voiced upon seeing the final product were implemented into the working prototype presented upon the conclusion of the hackathon. This mode of cooperation can be considered as a natural continuation of the model observed in the pre-design stage.

IV. DESIGN

*A. Platform architecture*

The F1 platform was designed as a client-server model and requires access to the Internet for the proper functioning. It might be a serious limitation for less developed and less populated countries, but as the platform is intended to be deployed in Poland, we decided to sacrifice versatility for simplicity.

Access to the system is possible either through a web site or a dedicated mobile application. Although both ways provide the same functionality, there are also substantial differences. Mobile application was designed to be most convenient for people offering support. Web site is more focused on posting requests for help. The reason for this differentiation is that mobile application will be more frequently used by younger volunteers and web browsers are better suited for older adults. In the case described in this paper senior participants were more familiar with traditional desktop or mobile computers than with smartphones. Moreover, computer web browsers can be more suitable for people with certain disabilities – e.g. visually impaired or with limited hand dexterity. On the other hand, mobile devices are becoming more and more widespread and the adoption of touch interface by the elderly can be faster and more effective than traditional computer interfaces (e.g. observed in our previous research studies). Thus, in the final product the application mode (web site or mobile app) is intended to be freely interchangeable at any time by the user.

### B. Functionality

The web-based application contains all key information and functions important for the user searching for the help of a volunteer. The screen informs the user about his or her previous favor requests, as well as the location of other users in the area. There is also an S.O.S button, which can be used in the case of emergency.

The mobile application view, used by the potential volunteer is shown in Figure1a. When using the mobile view, the user can view a map of the nearest area where favor requests are marked. When the user selects a favor, a brief description of the favor and the requesting user is provided.

### C. Security related features

Another direct benefit from applying participatory design approach refers to security issues. During the pre-design phase, older adults voiced their concerns related to the user security, taking into account threats for both parties of the process. The main areas in which they stated older adults require aid when it comes to security, were related to minimizing the risks of coming in contact with fake profiles, or malicious users, as well as dealing with problems of potential identity theft. The older adults involved in the project also stressed that the platform should be able to aid the older users while dealing with emergencies, when swift help is needed. These issues were addressed by applying the following solutions into the design of the platform.

*1) Trusted profiles:* Sign up is free for all and requires only a valid email account. Lowering the barrier makes the system more user-friendly but also prone to malicious users. Discussions with prospect users during the design phase revealed that the elderly are afraid of letting unknown people visit their apartments. To address this problem a voluntary procedure for confirming profiles was implemented. User profile might be confirmed by external organizations that are trustworthy: e.g. schools or local NGOs. Confirmation of the verified status is visible for everyone next to profile picture – see fig. 2.

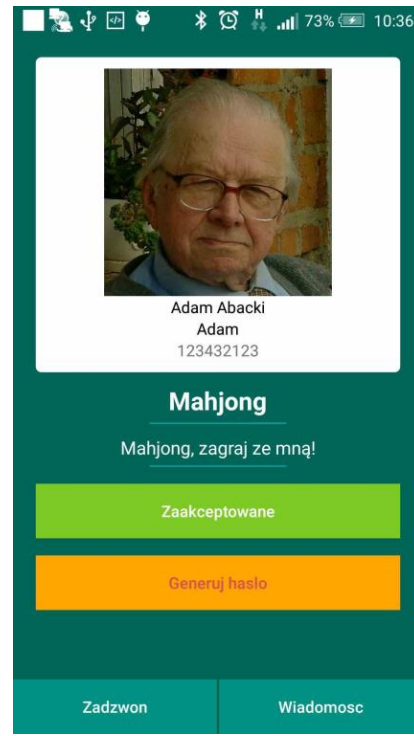

(a) Detailed information about request for assist.

Fig. 1: Mobile application dedicated to those who offer assistance.

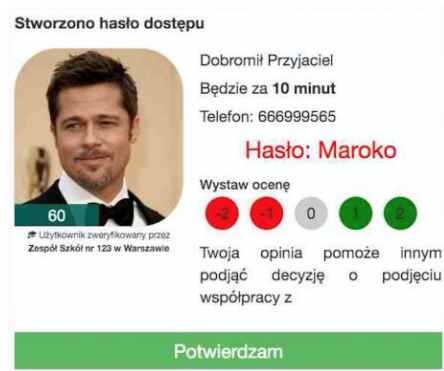

Fig. 2: Pop-up window displaying the confirmation details for selected volunteer

*2) Challenge-response authentication:* Next to the threat of fake or malicious profiles mentioned in previous section, yet another problem was identified by participatory approach. Even if identity of volunteer is confirmed on the platform, still we need a way to confirm it in real world, when a volunteer is knocking at the door of senior's apartment. This is the situation when digital system should face analogue world and a bottom-up approach proved to be helpful once again. Therefore, a standard challenge-response authentication has been adapted and implemented. The platform generates two keywords for both users. Elderly should ask about the right password before letting someone in. Passwords are randomly selected from a subset of Polish words to make them easy to remember and dictation by entry phone.

*3) Reputation score:* Limiting the amount of frauds is crucial for assuring the wide acceptance of the platform, but it is not enough. Next to deliberate and planed frauds, we can see poor quality service. Therefore, the platform contains a reputation management system. Every agreed and conducted service can be evaluated on Likert-type scale. To make the scale easier to understand for users, first two grades are red, neutral score is gray and the two positive levels are green. Sum of all evaluations for a given user are displayed next to the picture – see Fig. 2.

*4) Emergency button:* In real life exhaustive list of risks and threats is impossible to complete. Therefore, an emergency button has been added to the F1 platform.

## V. Conclusion and future work

In this paper we presented the report from the design and development of a platform for inter-generational exchange of favors, that was created by an inter-generational team during a 24-hour academic hackathon. The team working on the application consisted of younger programmers and older adults serving as representatives of the end users. Cooperation during pre-design and design stage helped to improve the quality of the final product, as well as helped to reduce the social distance between the members of the two generations. The platform was developed in a way that it addressed the issues raised by the end users represented here by two older adults. The younger programmers decided to follow the senior's idea of an application and implemented solutions that deal with the issues that older adults considered important while using an on-line platform. As a result, the programmers introduced a multistage system of security measures that foster the development of trust with a network of volunteers in social work.

In our future work we plan to further describe the live testing and implementation of the platform in real-life volunteer communities in Poland and abroad. We will also describe in more details the hackathon during which the platform was developed, and focus more on analyzing the modes of intergenerational cooperation observed during this event.

### Acknowledgments

This project has received funding from the European Union's Horizon 2020 research and innovation programme under the Marie Skłodowska-Curie grant agreement No 690962.

This project was partially supported by the infrastructure bought within the project „Heterogenous Computation Cloud" funded by the Regional Operational Programme of Mazovia Voivodeship.